\documentclass[aps,11pt]{revtex4}
\usepackage{amsmath}
\usepackage{bm}

\begin{document}
\title{Could we rotate proton decay away?}
\author{Ilja Dorsner$^{1}$}
\author{Pavel Fileviez P\'erez$^{1,2}$}
\affiliation{$^{1}$The Abdus Salam
International Centre for Theoretical Physics\\
Strada Costiera 11, 31014 Trieste, Italy. \\
$^{2}$Pontificia Universidad Cat\'olica de Chile \\
Facultad de F{\'\i}sica, Casilla 306 \\
Santiago 22, Chile.}
\begin{abstract}
In this work we investigate the possibility to completely rotate
away proton decay. We show that by choosing specific mass matrices
for fermions it is possible to accomplish this in flipped $SU(5)$.
\end{abstract}
\maketitle
Proton decay \cite{PatiSalam} is the most important prediction of
grand unified theories
~\cite{Georgi,SUSYSU(5),DeRujula,GeorgiMachacek,Barr}. For
phenomenological studies of its signatures see references
~\cite{Buras,Nathp1,Nathp2,Hisano,Raby,Goto,Dermisek}. It is
usually dominated by the gauge $d=6$ operators in any
non-supersymmetric grand unifying scenario. On the other hand, in
supersymmetric scenarios, proton decay is dominated by the $d=4$
and $d=5$~\cite{Dimopoulos,Weinberg,Zee,Sakai} operators. However,
the $d=4$ contributions can be forbidden by imposing the so-called
matter parity~\cite{Farrar} and there is always a way to suppress
the $d=5$ contributions~\cite{BabuBarr,Bajc1,Bajc2,Costa}.
Therefore, it looks as if the $d=6$ operators are the most
promising sources to test both supersymmetric and
non-supersymmetric scenarios. It is thus crucial if we can
establish how sensitive these operators to the parameters entering
in a grand unified theory are since they represent the source for
such an important signature of grand unification.

The idea of using the gauge $d=6$ dominated branching ratios for
the two-body nucleon decays to distinguish between different fermion
mass models has been around since the pioneering work of De
Rujula, Georgi and Glashow~\cite{DeRujula}. More recently, the
possibility to make a clear test of \textit{any}\/ grand
unified theory based on $SU(5)$ and $SO(10)$
with symmetric Yukawa couplings through the
decay channels into antineutrinos via $d=6$ gauge contributions
has been put forth~\cite{Pavel}. Similar program has also been
carried out in the context of flipped
$SU(5)$~\cite{DeRujula,GeorgiMachacek,Barr,Derendinger}. Namely,
the minimal flipped $SU(5)$~\cite{Antoniadis,Ellis1,Ellis2,Ellis3}
scenario can be tested by looking simultaneously at the decay
$p \rightarrow \pi^+ \overline{\nu}$ and the ratio $\tau(p \to K^0
e^+_{\alpha}) / \tau(p \to \pi^0 e^+_{\alpha})$~\cite{Dorsner}. It
is thus interesting to investigate how these conclusions change if
one departs from the flavor structure of the minimal
renormalizable theory.

It is well known that the gauge $d=6$ proton decay cannot be
rotated away, i.e., set to zero via particular choice of
parameters entering in a grand unified theory, in the framework of
conventional $SU(5)$~\cite{Georgi,SUSYSU(5)} theory with the
Standard Model particle content~\cite{Jarlskog,Mohapatra,Nandi}
. So, one might think that the
gauge $d=6$ operators and proton decay they govern are genuine
features of matter unification. In this work we show that this
might not be true. Namely, we demonstrate that it is possible to
\textit{completely}\/ rotate away the gauge $d=6$ contributions
for proton decay imposing simple conditions on fermion mixing. We
accomplish this in the framework of flipped $SU(5)$.

In order to appreciate all the difficulties involved in trying to
rotate proton decay away, let us first revisit the case of a
theory based on conventional $SU(5)$~\cite{Georgi,SUSYSU(5)}. In
the ordinary $SU(5)$ the integration of off-diagonal gauge bosons,
$V=(X,Y)=({{\bf 3}},{\bf 2},5/3)$, yields the following
operators~\cite{Weinberg,Zee,Sakai} in the physical
basis~\cite{Pavel}:
\begin{subequations}
\begin{eqnarray}
\label{4a} \textit{O}(e_{\alpha}^C, d_{\beta})_{SU(5)}&=&
c(e^C_{\alpha}, d_{\beta})_{SU(5)} \ \epsilon_{ijk} \
\overline{u^C_i} \ \gamma^{\mu} \ u_j \ \overline{e^C_{\alpha}} \
\gamma_{\mu} \ d_{k \beta}, \\
\label{4b} \textit{O}(e_{\alpha}, d^C_{\beta})_{SU(5)}&=&
c(e_{\alpha}, d^C_{\beta})_{SU(5)} \ \epsilon_{ijk} \
\overline{u^C_i} \ \gamma^{\mu} \ u_j \ \overline{d^C_{k \beta}} \
\gamma_{\mu} \ e_{\alpha},\\
\label{4c} \textit{O}(\nu_l, d_{\alpha}, d^C_{\beta} )_{SU(5)}&=&
c(\nu_l, d_{\alpha}, d^C_{\beta})_{SU(5)} \ \epsilon_{ijk} \
\overline{u^C_i} \ \gamma^{\mu} \ d_{j \alpha} \ \overline{d^C_{k
\beta}} \ \gamma_{\mu} \ \nu_l,\\
\label{4d} \textit{O}(\nu_l^C, d_{\alpha}, d^C_{\beta}
)_{SU(5)}&=& c(\nu_l^C, d_{\alpha}, d^C_{\beta})_{SU(5)} \
\epsilon_{ijk} \ \overline{d_{i \beta}^C} \ \gamma^{\mu} \ u_j \
\overline{\nu_l^C} \ \gamma_{\mu} \ d_{k \alpha},
\end{eqnarray}
\end{subequations}
where the coefficients that enter in the decay rate formulas take
the form (for the relevant decay formulas
see~\cite{Pavel,Dorsner}):
\begin{subequations}
\begin{eqnarray}
\label{5a} c(e^C_{\alpha}, d_{\beta})_{SU(5)}&=& k_1^2 \left[
V^{11}_1 V^{\alpha \beta}_2 + ( V_1 V_{UD})^{1
\beta}( V_2 V^{\dagger}_{UD})^{\alpha 1}\right], \\
\label{5b} c(e_{\alpha}, d_{\beta}^C)_{SU(5)} &=& k_1^2 V^{11}_1
V^{\beta \alpha}_3, \\
\label{5c} c(\nu_l, d_{\alpha}, d^C_{\beta})_{SU(5)}&=& k_1^2 (
V_1 V_{UD} )^{1 \alpha} ( V_3 V_{EN})^{\beta l}, \
\textrm{$\alpha=1$ or $\beta = 1$},\\
\label{5d} c(\nu_l^C, d_{\alpha}, d^C_{\beta})_{SU(5)}&=&0.
\end{eqnarray}
\end{subequations}
In the above expressions $k_1= g_5 M^{-1}_{V}$, where $M_{V} \sim
M_{GUT}\approx 10^{16}$\,GeV and $g_5$ are the masses of the
superheavy gauge bosons and the coupling at the GUT scale. $i$,
$j$ and $k$ are the color indices, $a$ and $b$ are the family
indices, and $\alpha, \beta =1,2$. The mixing matrices are: $V_1=
U_C^{\dagger} U$, $V_2=E_C^{\dagger}D$, $V_3=D_C^{\dagger}E$,
$V_{UD}=U^{\dagger}D$, and $V_{EN}=E^{\dagger}N$. Our convention
for the diagonalization of the up, down and charged lepton Yukawa
matrices is specified by $U^T_C Y_U U = Y_U^{\textrm{diag}}$, $
D^T_C Y_D D = Y_D^{\textrm{diag}}$, and $E^T_C Y_E E =
Y_E^{\textrm{diag}}$. The quark mixing is given by
$V_{UD}=U^{\dagger}D=K_1 V_{CKM} K_2$, where $K_1$ and $K_2$ are
diagonal matrices containing three and two phases, respectively.
$V_{CKM}$ is the Cabibbo-Kobayashi-Maskawa quark-mixing
matrix~\cite{Cabibbo,Kobayashi}.

The leptonic mixing $V_{EN}=K_3 V^D_l K_4$ in case of Dirac
neutrino, or $V_{EN}=K_3 V^M_l$ in the Majorana case. $V^D_l$ and
$V^M_l$ are the leptonic mixing matrices at low energy in the
Dirac and Majorana case, respectively.

We now show that the demand to rotate away proton decay leads
the conflict with the experimental data~\cite{Nandi}. In order to
set Eq.~\eqref{5b} to zero, the only possible choice is
$V_1^{11}=0$. [Setting $(V_3)^{\beta\alpha}$ to zero would violate
unitarity.] If we now look at Eq.~\eqref{5c}, there is only one
way to set to zero the coefficient entering in the decay channel
into antineutrinos. Namely, we have to choose $(V_1 V_{UD})^{1
\alpha}=0$. This, however, is not possible since it would imply
that, at least, $V_{CKM}^{13}$ is zero in conflict with the data.

Let us now investigate the same issue in flipped
$SU(5)$~\cite{DeRujula,GeorgiMachacek,Barr,Derendinger}. In this
case the gauge $d=6$ proton decay is mediated by
$V'=(X', Y')=({{\bf 3}},{\bf 2},-1/3)$. This time $d=6$ operators in
the physical basis are~\cite{Pavel}:
\begin{subequations}
\begin{eqnarray}
\label{6a} \textit{O}(e_{\alpha}^C, d_{\beta})_{SU(5)'}&=&
c(e^C_{\alpha}, d_{\beta})_{SU(5)'} \ \epsilon_{ijk} \
\overline{u^C_i} \ \gamma^{\mu} \ u_j \ \overline{e^C_{\alpha}} \
\gamma_{\mu} \ d_{k \beta}, \\
\label{6b} \textit{O}(e_{\alpha}, d^C_{\beta})_{SU(5)'}&=&
c(e_{\alpha}, d^C_{\beta})_{SU(5)'} \ \epsilon_{ijk} \
\overline{u^C_i} \ \gamma^{\mu} \ u_j \ \overline{d^C_{k \beta}} \
\gamma_{\mu} \ e_{\alpha},\\
\label{6c} \textit{O}(\nu_l, d_{\alpha}, d^C_{\beta} )_{SU(5)'}&=&
c(\nu_l, d_{\alpha}, d^C_{\beta})_{SU(5)'} \ \epsilon_{ijk} \
\overline{u^C_i} \ \gamma^{\mu} \ d_{j \alpha}
\ \overline{d^C_{k \beta}} \ \gamma_{\mu} \ \nu_l, \\
\label{6d} \textit{O}(\nu_l^C, d_{\alpha}, d^C_{\beta}
)_{SU(5)'}&=& c(\nu_l^C, d_{\alpha}, d^C_{\beta})_{SU(5)'} \
\epsilon_{ijk} \ \overline{d_{i \beta}^C} \ \gamma^{\mu} \ u_j \
\overline{\nu_l^C} \ \gamma_{\mu} \ d_{k \alpha},
\end{eqnarray}
\end{subequations}
where
\begin{subequations}
\begin{eqnarray}
\label{7a} c(e^C_{\alpha}, d_{\beta})_{SU(5)'}&=&0,\\
 \label{7b}
c(e_{\alpha}, d_{\beta}^C)_{SU(5)'} &=&
k_2^2 (V_4 V^{\dagger}_{UD} )^{\beta 1} ( V_1 V_{UD} V_4^{\dagger} V_3)^{1 \alpha},\\
\label{7c} c(\nu_l, d_{\alpha}, d^C_{\beta})_{SU(5)'}&=&  k_2^2
V_4^{\beta \alpha}( V_1 V_{UD} V^{\dagger}_4 V_3 V_{EN})^{1l}, \
\textrm{$\alpha=1$ or $\beta = 1$}, \\
\label{7d} c(\nu_l^C, d_{\alpha}, d^C_{\beta})_{SU(5)'}&=& k_2^2
\left[ ( V_4 V^{\dagger}_{UD} )^{\beta
 1} ( U^{\dagger}_{EN} V_2)^{l \alpha }+ V^{\beta \alpha}_4
 (U^{\dagger}_{EN} V_2 V^{\dagger}_{UD})^{l1}\right], \
\textrm{$\alpha=1$ or $\beta = 1$}.
\end{eqnarray}
\end{subequations}
Notice that we use the subscripts $SU(5)'$ for flipped $SU(5)$. In
the above equations, the mixing matrices $V_4=D_C^{\dagger} D$,
and $U_{EN}= E_C^{\dagger} N_C$. The factor $k_2= g_5'/ M_{V'}$,
where $g_5'$ is given by the unification of $\alpha_2$ and
$\alpha_3$.

Let us see if it is possible to rotate away the proton decay in
flipped $SU(5)$. To set Eq.~\eqref{7c} to zero, we can only choose
$V_4^{\beta\alpha}= (D_C^{\dagger} D)^{\beta \alpha}=0$, where
$\alpha=1$ or $\beta=1$. We could think about possibility of
making both Eqs.~\eqref{7b} and \eqref{7d} zero, choosing $(V_4
V_{UD}^\dagger)^{\beta 1}=0$, however, this is in contradiction
with the measurements of the CKM angles. Since in flipped $SU(5)$
the neutrino is Majorana, we only have to suppress Eq.~\eqref{7b}.
This can be accomplished by setting
$(V_1 V_{UD} V_4^{\dagger} V_3)^{1 \alpha}=(U_C^{\dagger} E)^{1
  \alpha}=0$.
Notice that this is completely unrelated to our condition on
$V_4$. Thus, there is no contradiction with unitarity constrains
nor conflict with any experimental measurements of mixing angles.
As you can appreciate, in the context of flipped $SU(5)$, it is
possible to \textit{completely} rotate away the gauge $d=6$
contributions in a consistent way, if we impose these two
conditions at 1\,GeV.

We stress that in minimal renormalizable flipped
$SU(5)$~\cite{Antoniadis,Ellis1,Ellis2,Ellis3} it is not possible
to satisfied the first condition, since $Y_D=Y_D^T$ implies
$V_4=K^*_d$, where $K_d$ is a diagonal matrix containing three CP
violating phases. However, as we know, in general we have to take
into account the nonrenormalizable operators, which are very
important for fermion masses and which invariably lead to
modification of naive predictions. Therefore in general, in the
context of flipped $SU(5)$, we are allowed to impose our
conditions and remove the gauge operators for proton decay.

Note that the main difference between the $SU(5)$ analysis and
flipped $SU(5)$ one is that the unitary constraint that prevents
us to rotate away proton decay in conventional $SU(5)$ does not
operate in the latter case. In other words, the coefficients which
depend on $\alpha$ and $\beta$ with $\alpha=1$ or $\beta =1$ have
different impact in those two scenarios (see Eqs.~\eqref{5b} and
\eqref{7c}).

What these two conditions that remove $d=6$ operators imply for
the structure of the fermion sector? We give one example. Let us
choose the basis where the up quark mass matrix is diagonal. In
this case we have:
\begin{subequations}
\begin{eqnarray}
Y_D &=& K^*_1 \ V_{CKM}^* \ K_2^* \ A \ Y_D^{\textrm{diag}} \
K_2^* \ V_{CKM}^{\dagger} \ K_1^*,
\\
Y_E &=& E_C^* \ Y_E^{\textrm{diag}} \ E^{\dagger},
\\
Y_N &=& K_3 ^* \ V_l^* \ E^* \ Y_N^{\textrm{diag}} \ E^{\dagger} \
V_l^{\dagger} \ K_3^{*},
\end{eqnarray}
\end{subequations}
where $|E^{13}|=1$ and $A$ is a unitary matrix, with
$|A^{13}|=|A^{22}|=|A^{31}|=1$.

In order to understand if it is possible to suppress all
contributions to proton decay we assume the matter parity to be an
exact symmetry and proceed with the analysis of the Higgs $d=6$
and $d=5$ contributions. In SUSY flipped $SU(5)$ the interactions
for triplets are given by:
\begin{equation}
W_T = \int \ d^2\theta \ \left[\hat{Q} \ \underline{A} \ \hat{Q} \
\hat{T} \ + \ \hat{D^C} \ \underline{B} \ \hat{N^C} \ \hat{T} \ +
\ \hat{Q} \ \underline{C} \ \hat{L} \ \hat{\overline{T}} \ + \
\hat{D^C} \ \underline{D} \ \hat{U^C} \
\hat{\overline{T}}\right]+\textrm{h.c.},
\end{equation}
where the matrices $\underline{A}$, $\underline{B}$,
$\underline{C}$, and $\underline{D}$ are:
\begin{subequations}
\begin{eqnarray}
\underline{A}&=& Y^{\textrm{ren}}_D \ + \frac{M_{GUT}}{M_{Planck}}
\ Y_1+ \frac{M_{GUT}^2}{M_{Planck}^2} \ Y_2 ,
\\
\underline{B}&=& Y^{\textrm{ren}}_D \ + \frac{M_{GUT}}{M_{Planck}}
\ Y_3+ \frac{M_{GUT}^2}{M_{Planck}^2} \ Y_4 ,
\\
\underline{C}&=& Y^{\textrm{ren}}_U \ + \frac{M_{GUT}}{M_{Planck}}
\ Y_5+ \frac{M_{GUT}^2}{M_{Planck}^2} \ Y_6 ,
\\
\underline{D}&=& Y^{\textrm{ren}}_U \ + \frac{M_{GUT}}{M_{Planck}}
\ Y_7+  \frac{M_{GUT}^2}{M_{Planck}^2}\ Y_8,
\end{eqnarray}
\end{subequations}
up to the second order in $M_{GUT}/ M_{Planck}$ expansion.
$Y^{\textrm{ren}}_D$ and $Y^{\textrm{ren}}_U$ are the Yukawa
matrices at the renormalizable level for down and up quarks,
respectively. $Y_i$, $i=1..8$, are the contributions coming from
the non-renormalizable terms.

Now, notice that we could forbid the $d=5$ and Higgs $d=6$
contributions, imposing
the conditions $\underline{A}^T = - \underline{A}$
and $(D_C^T \ \underline{D} \ U_C)^{ij}=0$, except for $i = j =3$.
For similar approach see~\cite{Dvali,Berezhiani}.
This confirms that it is possible to
completely rotate proton decay away in flipped $SU(5)$ context.

In this work we review the possibilities to suppress all operators
for proton decay. We revisit conventional $SU(5)$ to show that in
this scenario it is not possible to rotate away proton decay. We
further investigate the case of flipped $SU(5)$ finding that there
it is possible to \textit{completely}\/ eliminate the gauge $d=6$
operators. In the same context we show the way to remove the Higgs
$d=6$ and $d=5$ contributions. Our main result---the possibility
to rotate away proton decay in flipped $SU(5)$---shows the lack of
robustness of the gauge $d=6$ contributions under departure from
the ``naive" assumptions for the parameters entering matter
unifying theories.
\begin{acknowledgments}
P.F.P thanks the High Energy Section of the ICTP for their
hospitality and support. This work was supported in part
by CONICYT/FONDECYT under contract $N^{\underline 0} \ 3050068$.
\end{acknowledgments}


\begin{thebibliography}{99}
\bibitem{PatiSalam}
J.~C.~Pati and A.~Salam,
Phys.\ Rev.\ D {\bf 8} (1973) 1240; Phys.\ Rev.\ Lett.\  {\bf 31}
(1973) 661; Phys.\ Rev.\ D {\bf 10} (1974) 275.
\bibitem{Georgi}
H.~Georgi and S.~L.~Glashow, Phys.\ Rev.\ Lett.\  {\bf 32}, 438
(1974).
\bibitem{SUSYSU(5)}
S.~Dimopoulos and H.~Georgi, Nucl.\ Phys.\ B {\bf 193} (1981) 150.
\bibitem{DeRujula}
A.~De Rujula, H.~Georgi and S.~L.~Glashow, Phys.\ Rev.\ Lett.\
{\bf 45} (1980) 413.
\bibitem{GeorgiMachacek}
H.~Georgi, S.~L.~Glashow and M.~Machacek, Phys.\ Rev.\ D {\bf 23}
(1981) 783.
\bibitem{Barr}
S.~M.~Barr, Phys.\ Lett.\ B {\bf 112}, 219 (1982).
\bibitem{Buras}A.~J.~Buras, J.~R.~Ellis, M.~K.~Gaillard and
D.~V.~Nanopoulos, Nucl.\ Phys.\ B {\bf 135} (1978)66.
\bibitem{Nathp1} P.~Nath, A.~H.~Chamseddine and R.~Arnowitt, Phys.\ Rev.\ D
{\bf 32} (1985) 2348.
\bibitem{Nathp2}
R.~Arnowitt, A.~H.~Chamseddine and P.~Nath, Phys.\ Lett.\ B {\bf
156} (1985) 215.
\bibitem{Hisano}
J.~Hisano, H.~Murayama and T.~Yanagida, Nucl.\ Phys.\ B {\bf 402}
(1993) 46.
\bibitem{Raby}
V.~Lucas and S.~Raby, Phys.\ Rev.\ D {\bf 55} (1997) 6986.
\bibitem{Goto} T.~Goto and T.~Nihei, Phys.\ Rev.\ D {\bf 59} (1999)
115009.
\bibitem{Dermisek}
R.~Dermisek, A.~Mafi and S.~Raby, Phys.\ Rev.\ D {\bf 63} (2001)
035001.
\bibitem{Dimopoulos}
S.~Dimopoulos, S.~Raby and F.~Wilczek, Phys.\ Lett.\ B {\bf 112}
(1982) 133.
\bibitem{Weinberg}
S.~Weinberg, Phys.\ Rev.\ Lett.\ {\bf 43} (1979) 1566;
Phys.\ Rev.\ D {\bf 22} (1980) 1694; Phys.\ Rev.\ D {\bf 26} (1982) 287.
\bibitem{Zee}F.~Wilczek and A.~Zee, Phys.\ Rev.\ Lett.\  {\bf
43} (1979) 1571.
\bibitem{Sakai}
N.~Sakai and T.~Yanagida, Nucl.\ Phys.\ B {\bf 197} (1982) 533.
\bibitem{Farrar}
G.~R.~Farrar and P.~Fayet, Phys.\ Lett.\ B {\bf 76} (1978) 575.
\bibitem{BabuBarr}
K.~S.~Babu and S.~M.~Barr, Phys.\ Rev.\ D {\bf 48} (1993) 5354.
\bibitem{Bajc1} B.~Bajc, P.~Fileviez~P\'erez and G.~Senjanovi\'c,
Phys.\ Rev.\ D {\bf 66} (2002) 075005.
\bibitem{Bajc2} B.~Bajc, P.~Fileviez~P\'erez and G.~Senjanovi\'c,
hep-ph/0210374.
\bibitem{Costa} D.~Emmanuel-Costa and S.~Wiesenfeldt, Nucl.\ Phys.\ B {\bf 661} (2003) 62.
\bibitem{Pavel}
P.~Fileviez~P\'erez, Phys.\ Lett.\ B {\bf 595} (2004) 476.
\bibitem{Derendinger}
J.~P.~Derendinger, J.~E.~Kim and D.~V.~Nanopoulos, Phys.\ Lett.\ B
{\bf 139}, 170 (1984).
\bibitem{Antoniadis}
I.~Antoniadis, J.~R.~Ellis, J.~S.~Hagelin and D.~V.~Nanopoulos,
Phys.\ Lett.\ B {\bf 194}, 231 (1987).
\bibitem{Ellis1}
J.~R.~Ellis, J.~S.~Hagelin, S.~Kelley and D.~V.~Nanopoulos, Nucl.\
Phys.\ B {\bf 311} (1988) 1.
\bibitem{Ellis2}
J.~R.~Ellis, J.~L.~Lopez, D.~V.~Nanopoulos and K.~A.~Olive, Phys.\
Lett.\ B {\bf 308} (1993) 70.
\bibitem{Ellis3}
J.~R.~Ellis, D.~V.~Nanopoulos and J.~Walker, Phys.\ Lett.\ B {\bf
550} (2002) 99.
\bibitem{Dorsner}
I.~Dorsner and P.~Fileviez~P\'erez, arXiv:hep-ph/0409095.
\bibitem{Jarlskog}
C.~Jarlskog, Phys.\ Lett.\ B {\bf 82} (1979) 401.
\bibitem{Mohapatra}
R.~N.~Mohapatra, Phys.\ Rev.\ Lett.\  {\bf 43} (1979) 893.
\bibitem{Nandi}
S.~Nandi, A.~Stern and E.~C.~G.~Sudarshan, Phys.\ Lett.\ B {\bf
113} (1982) 165.
\bibitem{Cabibbo}
N.~Cabibbo, Phys.\ Rev.\ Lett.\  {\bf 10} (1963) 531.
\bibitem{Kobayashi}
M.~Kobayashi and T.~Maskawa, Prog.\ Theor.\ Phys.\  {\bf 49}
(1973) 652.
\bibitem{Dvali}
G.~R.~Dvali, Phys.\ Lett.\ B {\bf 287} (1992) 101.
\bibitem{Berezhiani}
Z.~Berezhiani, arXiv:hep-ph/9602325.
\end{thebibliography}
\end{document}